# L'auto-efficacité numérique comme fondement d'un cadre d'usage des IA génératives dans les pratiques professionnelles des enseignants-chercheurs

Fatiha TALI OTMANI[1,2]

[1]IUniversité Toulouse Jean Jaurès, UMR-EFTS
[2]INSEI, Grhapes

**Résumé.** Cette recherche explore le rôle de l'auto-efficacité numérique dans l'appropriation des intelligences artificielles génératives (IAG) par les enseignants-chercheurs. S'appuyant sur la théorie sociocognitive de Bandura et le concept de cadre d'usage de Flichy, notre étude examine les relations entre les niveaux d'auto-efficacité numérique et les profils d'utilisation des IAG. Une enquête menée auprès de 265 enseignants-chercheurs a permis d'identifier trois profils d'utilisateurs (Investis, Réservés réflexifs, Réfractaires critiques) et de valider une échelle d'auto-efficacité numérique à trois dimensions. Les résultats révèlent une association significative entre les profils d'auto-efficacité et les modalités d'appropriation des IAG. Sur cette base, nous proposons un cadre d'usage différencié, intégrant quatre configurations sociotechniques, des trajectoires d'appropriation adaptées aux profils d'auto-efficacité, et des dispositifs institutionnels d'accompagnement personnalisés.
**Mots-clés :** auto-efficacité, IA génératives, cadre d'usage, enseignants-chercheurs, transformations, pratiques professionnelles

*Abstract.* This research explores the role of digital self-efficacy in the appropriation of generative artificial intelligence (GAI) by higher education faculty. Drawing on Bandura's sociocognitive theory and Flichy's concept of usage framework, our study examines the relationships between levels of digital self-efficacy and GAI usage profiles. A survey of 265 faculty members identified three user profiles (Engaged, Reflective Reserved, Critical Resisters) and validated a three-dimensional digital self-efficacy scale. Results reveal a significant association between self-efficacy profiles and GAI appropriation patterns. Based on these findings, we propose a differentiated usage framework integrating four sociotechnical configurations, appropriation trajectories adapted to self-efficacy profiles, and personalized institutional support mechanisms.
.

# 1. INTRODUCTION

L'émergence des intelligences artificielles génératives (IAG) constitue une rupture paradigmatique dans l'écosystème académique, bouleversant les pratiques professionnelles des enseignants-chercheurs. Ces systèmes, capables de produire des contenus sophistiqués à partir de données préexistantes, remettent en question les dispositifs d'enseignement et d'évaluation traditionnels (Perkins *et al.*, 2024). Face à ces transformations rapides, les enseignants-chercheurs (EC) développent des stratégies d'appropriation ou de résistance qui reflètent la complexité des rapports entre technologie et identité professionnelle.

La théorie sociocognitive de Bandura (2019) offre un cadre pertinent pour analyser ces dynamiques d'adaptation. L'auto-efficacité, définie comme la croyance d'un individu en sa capacité à mobiliser les ressources nécessaires pour accomplir une tâche donnée, apparaît comme un médiateur cardinal de l'agentivité. Cette dernière renvoie à l'aptitude à exercer intentionnellement une influence sur ses propres conduites dans un contexte spécifique (Jézégou, 2019). Ce concept s'avère particulièrement fécond pour appréhender les processus d'appropriation des IAG en milieu académique (Nikolopoulou, 2025).

Notre recherche vise à conceptualiser un cadre d'usage des IAG qui articule dimensions individuelles et institutionnelles pour favoriser leur intégration raisonnée. Notre étude explore deux objectifs : d'une part, mieux cerner les liens entre niveau d'auto-efficacité numérique et mécanismes d'appropriation ou de résistance face aux IAG chez les enseignants-chercheurs et d'autre part, modéliser un cadre d'usage adapté aux différents profils d'utilisateurs dans l'enseignement supérieur en fonction de leur niveau d'auto-efficacité numérique.

La première partie de cet article présente le cadre théorique permettant de définir les IAG et les enjeux entourant leurs usages, de circonscrire le concept de cadre d'usage, puis de saisir les concepts d'auto-efficacité et d'agentivité dans l'approche sociocognitive en les articulant à la notion de cadre d'usage des technologies éducatives. La deuxième partie expose la méthodologie adoptée, suivie d'une troisième partie détaillant les résultats de notre enquête. La quatrième partie propose un cadre d'usage fondé sur l'auto-efficacité, avant de discuter, en conclusion, les implications théoriques et pratiques de cette recherche.

# 2. CADRE THÉORIQUE

## 2.1. LES IA GENERATIVES DANS L'ENSEIGNEMENT SUPERIEUR

Dans le contexte de l'enseignement supérieur, de nombreux systèmes d'IA sont mobilisés selon les pratiques (de recherche, d'enseignement, d'évaluation…) et il est nécessaire de les définir et en saisir les enjeux dans l'optique de formaliser un cadre d'usage.

### 2.1.1. Définitions des systèmes d'IA

Tout comme pour le numérique, une multiplicité de définitions co-existe pour l'IA. Russell et Norvig (2021, p. 1) définissent l'IA comme « l'étude des agents [intelligents] qui perçoivent leur environnement et entreprennent des actions. Chacun de ces agents est implémenté par une fonction qui établit une correspondance entre les perceptions et les actions, et nous examinons différentes manières de représenter ces fonctions, telles que les systèmes de production, les agents réactifs, les planificateurs logiques, les réseaux neuronaux et les systèmes décisionnels probabilistes »[1].

L'IA générative (IAG) constitue une branche spécifique de l'IA, caractérisée par sa capacité à produire du contenu en s'appuyant sur des modèles d'apprentissage complexes (Class et De la Higuera, 2024). Les systèmes d'IAG ont pris une place de plus en plus importante dans le paysage académique depuis leur accès démocratisé en 2022.

Selon la définition de l'UNESCO, l'IAG est une technologie capable de « générer automatiquement du contenu en réponse à des messages rédigés dans des interfaces de conversation en langage naturel » (Holmes et Miao, 2024, p. 8). Elle s'inscrit dans un continuum d'innovations en IA, aux côtés d'autres systèmes capables de générer des prédictions, des classifications ou des recommandations (Nashed *et al.*, 2022). Les modèles d'IAG, tels que ChatGPT, Claude ou Mistral, utilisent des architectures de réseaux neuronaux profonds qui analysent les patterns statistiques de vastes ensembles de données pour générer des réponses contextualisées (Holmes et Miao, 2024). Au regard de nos travaux, l'IAG peut être vue comme un outil médiateur : 1. Dans le processus d'enseignement et d'apprentissage, reliant l'enseignant à la fois au savoir et aux étudiants ; 2. Dans le processus de recherche, mettant en lien savoirs, données générées ou analysées, productions d'écrits scientifiques, tout en étant façonnée par le contexte sociocognitif et institutionnel dans lequel elle est utilisée.

### 2.1.2. Différents enjeux d'intégration de l'IAG dans l'enseignement supérieur

Les EC articulent à la fois des pratiques liées à la recherche mais également à l'enseignement, l'évaluation ou à des charges administratives. La présence croissante des IAG dans l'environnement académique n'est pas sans conséquences. Ainsi, si les cadres réglementaires récents, comme l'*AI Act* européen (2024) ou la Convention-cadre du Conseil de l'Europe sur l'IA (2024), visent à encadrer les usages de l'IA à travers une régulation par les risques, la littérature scientifique montre que ces dispositifs doivent être complétés par des analyses éthiques spécifiques au contexte éducatif. En effet, l'IA soulève des enjeux éthiques dans l'enseignement supérieur sous trois dimensions principales : la conception, les données et l'usage (Collin et Marceau, 2022).

De nombreux biais ont été mis à jour quant au recours aux IAG dans les pratiques de recherche et les pratiques enseignantes des EC (Zollinger, 2024). Ces systèmes computationnels sont entraînés sur des données sélectionnées par les programmateurs et les propriétaires de ces outils, impliquant de fait des effets de sur-représentation de certaines données issues des pays occidentaux et de sous-représentation de celles provenant des cultures de pays du Sud (Holmes et Miao, 2024), de certains types de populations (personnes en situation de handicap par exemple, ou encore des biais de genre).

Tout d'abord, il est à noter que l'*AI Act*[2], adopté en 2024, a introduit une approche de régulation fondée sur les risques, classant les systèmes d'IA selon leur potentiel de nuisance

---

[1] Notre traduction

[2] Loi sur l'intelligence artificielle (règlement (UE) 2024/1689), version du Journal officiel du 13 juin 2024 visible ici : https://artificialintelligenceact.eu/fr/l-acte/. Depuis le 2 février 2025, les systèmes d'IA jugés à « risque inacceptable » sont interdits en Europe. Les usages interdits recouvrent : la notation sociale, la manipulation cognitive, la reconnaissance biométrique en temps réel et la prédiction criminelle basée sur

pour les droits fondamentaux. Dans le domaine éducatif, les systèmes de tutorat intelligent, les outils d'évaluation automatisée et les plateformes de surveillance des examens sont explicitement désignés comme « à haut risque ». Cette classification impose des obligations spécifiques, notamment en termes de transparence, de robustesse et de non-discrimination. Ces exigences s'alignent sur les principes éthiques formulés par (Floridi, 2013) dans son analyse de l'éthique de l'information, où la transparence et la justice algorithmique sont identifiées comme des conditions fondamentales pour garantir l'équité des systèmes automatisés. Cependant, l'application de ces principes dans l'enseignement supérieur reste limitée par une documentation encore embryonnaire des risques éthiques spécifiques (Zawacki-Richter et al., 2019).

Dans le domaine de la recherche, l'*AI Act* prévoit une exception pour préserver l'innovation : les obligations ne s'appliquent pas « aux systèmes et modèles d'IA, y compris leurs résultats, spécifiquement développés et mis en service aux seules fins de la recherche et du développement scientifique, ni à leurs sorties » (Frayret *et al*., 2024, p. 3). Cependant, le périmètre exact de cette exception reste flou. Si les modèles d'IA développés et utilisés dans le cadre d'un projet de recherche sont clairement concernés, le statut des IA conçues en appui à la recherche est moins évident. En revanche, les IA utilisées pour la recherche mais conçues à d'autres fins (comme un « modèle d'IAG commercial dans le cadre d'un projet de recherche ») seront soumises à l'*AI Act* (Frayret et al., 2024, p. 3).

Ces différents points sont donc à prendre en compte dans l'élaboration d'un cadre d'usage respectueux du volet éthique.

## 2.2. VERS UNE CONCEPTUALISATION DES CADRES D'USAGE DES TECHNOLOGIES EDUCATIVES

### 2.2.1. Définition et composantes d'un cadre d'usage

Le concept de cadre d'usage, développé par Flichy (2008), offre une approche théorique particulièrement intéressante pour analyser l'appropriation des technologies numériques dans l'enseignement supérieur. Contrairement aux approches techno-centrées, ce concept s'intéresse à la dimension sociale des usages technologiques et à leur construction collective. L'auteur (2008, p. 164-165) définit le cadre d'usage comme « celui qui décrit le type d'activités sociales proposées par la technique, qui la positionne dans l'éventail des pratiques sociales, des routines de la vie quotidienne, et précise les publics envisagés, les lieux et les situations où cette technique peut se déployer ».

Cette définition met en évidence la double fonction du cadre d'usage : cognitive et symbolique d'une part, permettant aux acteurs de donner du sens à la technologie ; organisationnelle d'autre part, structurant les interactions des utilisateurs avec l'objet technique et entre eux. Dans le contexte des IAG dans l'enseignement supérieur, le cadre d'usage détermine comment ces outils sont perçus (comme assistants à la production scientifique, menaces pour l'intégrité académique, ou instruments pédagogiques innovants), ainsi que les situations où leur utilisation est jugée appropriée.

---

l'apparence. Pour les entreprises qui n'appliquent pas la loi, sont prévues des amendes allant jusqu'à 35 millions d'euros ou 7 % de leur chiffre d'affaires annuel. À noter que les forces de l'ordre ont quelques exceptions d'usages. Une mise à jour en juillet 2025 pose le cadre du droit d'auteur également : https://digital-strategy.ec.europa.eu/en/policies/contents-code-gpai

Une caractéristique essentielle du *cadre d'usage* est qu'il résulte d'une construction collective impliquant de multiples acteurs. Comme le souligne Flichy (2008, p. 165), « la définition du cadre d'usage n'est pas assurée par l'usager. Elle est le résultat de l'action conjointe de tous les acteurs de l'activité technique, le concepteur, les nombreux médiateurs [...], mais aussi l'usager ». Cette perspective permet de dépasser la dichotomie traditionnelle entre concepteurs et utilisateurs pour envisager un processus plus complexe de co-construction des usages, ce que nous proposons ici à partir des pratiques déclarées des EC et de cadres d'usage existants.

L'articulation entre stabilité et flexibilité constitue une autre caractéristique fondamentale du cadre d'usage. Flichy (2008, p. 166) précise que « ce cadre d'usage ne définit qu'un ensemble d'usages standard autour desquels les usagers peuvent broder ». Cette conception permet de comprendre comment un cadre d'usage institutionnalisé des IAG peut co-exister avec des appropriations variées selon les disciplines et les profils d'EC, dont les profils liés à leur niveau d'auto-efficacité numérique comme nous allons le décliner.

Le cadre d'usage se distingue des modèles d'adoption technologique dominants comme le Technology Acceptance Model (TAM) de Davis (1989) ou la Unified Theory of Acceptance and Use of Technology (UTAUT) de Venkatesh (2022). Ces modèles privilégient une approche individuelle centrée sur les déterminants de l'intention d'usage, tandis que le cadre d'usage s'intéresse davantage aux processus collectifs de construction de sens et aux pratiques effectives.

### 2.2.2. Des cadres d'usage pour les technologies éducatives ?

Dans le champ de l'éducation, l'usage des technologies numériques par les enseignants s'inscrit dans des cadres définis et régulés par les institutions éducatives. Toutefois, ces cadres institutionnels peuvent entrer en tension avec les modalités d'appropriation que les enseignants construisent eux-mêmes, ainsi qu'avec les injonctions prescrites par les politiques éducatives (Bouakka et Chakouk, 2024). Ces tensions se traduisent souvent par des formes de résistance, qui freinent l'intégration effective des outils numériques et limitent leur mobilisation au sein des pratiques pédagogiques (Dias-Chiaruttini *et al.*, 2020), d'autant plus que le recours aux IA ne peut être fait sans une analyse critique des effets sur l'éducation ( Holmes *et al*., 2025)

Différentes instances ont publié des documents encadrant l'usage des IA dans l'éducation. En France, le ministère de l'Éducation nationale et de la Jeunesse a promulgué en 2025 un cadre réglementaire définissant les modalités d'utilisation de l'IA au sein des établissements d'enseignement. Ce cadre autorise l'usage de l'IA à condition qu'il respecte les valeurs républicaines, la protection des données personnelles et une approche éthique. Il structure l'utilisation pédagogique de l'IA par niveau scolaire, avec une introduction progressive : sensibilisation dès le primaire, manipulation encadrée à partir de la 4$^e$, et usage plus autonome au lycée. Puis, dans le contexte québécois, le Ministère de l'Enseignement supérieur du Québec a également émis en 2025 un document nommé « Intégration responsable de l'intelligence artificielle dans les établissements d'enseignement supérieur ». Il propose des outils pratiques comme une taxonomie des usages, une grille d'autodiagnostic et des exemples de structures de gouvernance, tout en insistant sur l'importance d'aligner les choix en IA avec les valeurs institutionnelles. Publié en 2025, le *Ethical & Legal Guide on the Responsible Integration of AI in Education* (Bertel et al., 2025) constitue le premier cadre européen explicitement destiné aux établissements scolaires et d'enseignement supérieur. À la différence de nombreux documents sectoriels, il articule prescriptions éthiques (respect de l'autonomie, équité, supervision humaine) et obligations juridiques (*AI Act*, RGPD), en opérant une triade « avant-pendant-après » qui balise l'ensemble du cycle d'usage.

Au niveau des établissements d'enseignement supérieur, plusieurs approches se distinguent : les unes très prescriptives, d'autres fondées sur des principes mis en exergue, et d'autres articulent de manière hybride les deux. Par exemple, l'Université de Montréal (2024) a développé des « Lignes directrices pour une utilisation appropriée de l'intelligence artificielle générative aux études supérieures » qui mettent l'accent sur l'intégrité académique et définissent trois niveaux d'utilisation : interdite, autorisée avec déclaration, et encouragée. De son côté, l'Université Laval (2024) a publié des « Principes directeurs concernant l'intelligence artificielle dans l'enseignement et l'apprentissage » qui proposent un cadre souple fondé sur cinq principes : transparence, équité, respect de l'autonomie des apprenants, intégrité et responsabilité. Au-delà, diverses chartes ont été publiées par des universités en France (comme celle de l'université d'Orléans (Cormaty, 2025) et dans le monde comme celle de l'université Stanford (2023) qui a opté pour une « charte de l'IA en éducation » qui promeut une « approche adaptative » permettant aux facultés de développer leurs propres politiques d'usage en fonction de leurs disciplines spécifiques, tout en respectant un ensemble de principes communs. Par ailleurs, diverses organisations internationales ont proposé des cadres structurants comme l'UNESCO (2023, 2025) ou l'OCDE (2023).

Les points communs des cadrages proposés par les différentes institutions se retrouvent sur la nécessité de la formation des enseignants à l'usage des IA, d'un développement de la littératie de l'IA des étudiants ainsi qu'une sensibilisation aux enjeux éthiques et sociaux. Différents contextes d'utilisation sont également relevés : pour l'enseignement et l'apprentissage ; pour l'évaluation ; pour la recherche ; pour le volet administratif. Chaque contexte appelle des règles spécifiques, comme le souligne le cadre québécois.

Cependant, au sens de Flichy (2008), un cadre d'usage décrit les activités sociales proposées par une technologie, leurs publics, ainsi que les lieux et situations d'appropriation. Les dispositifs nationaux (MENJ, 2025 ; MES, 2025) répondent pleinement à cette définition, en prescrivant des usages différenciés selon les contextes éducatifs sans pour autant intégrer suffisamment la dimension épistémique, essentielle dans l'appropriation des IAG en contexte académique. Les lignes directrices institutionnelles (UdeM, 2024 ; ULaval, 2024 ; Stanford, 2023) s'en rapprochent également, mais à une échelle locale, avec des prescriptions contextualisées pour leurs publics respectifs. A contrario, les publications émises par l'UNESCO (2023) et l'OCDE (2023) n'entrent pas dans cette définition. Ils relèvent plus particulièrement de cadres de principes et non de cadres d'usage au sens de Flichy (2008). Le cadre d'usage européen (Bertel *et al.*, 2025) relève partiellement de cette notion : il fournit bien un référentiel normatif qui oriente les pratiques, formalise les rôles (provider/deployer) et anticipe les représentations (IA fiable, humaine, durable). Toutefois, il reste essentiellement réglementaire et prescriptif ; il n'intègre ni l'analyse des usages réels, ni la dimension de négociation sociale chère à Flichy.

Par ailleurs, les travaux actuels n'explorent pas suffisamment les facteurs individuels qui influencent la participation des acteurs à la construction et à l'activation des cadres d'usage, en particulier les facteurs psychosociaux qui pourraient expliquer les variations dans les usages. Pour combler cette lacune, nous proposons d'articuler le concept de cadre d'usage avec celui d'auto-efficacité (AE), qui permet d'éclairer les différences individuelles dans la capacité à s'approprier les technologies et à contribuer à la définition de leurs usages légitimes.

## 2.3. L'AUTO-EFFICACITE NUMERIQUE COMME CONCEPT MEDIATEUR

L'auto-efficacité (AE), concept central de la théorie sociocognitive de Bandura (2019), apparaît comme un médiateur potentiel particulièrement pertinent pour comprendre comment les EC participent à la construction et à l'activation des cadres d'usage des IAG. Définie comme « la croyance de l'individu en sa capacité d'organiser et d'exécuter la ligne de conduite requise pour produire des résultats souhaités » (Bandura, 2007, p. 12), l'auto-efficacité influence significativement l'engagement des individus dans de nouvelles pratiques et leur persévérance face aux obstacles. Les quatre sources de l'auto-efficacité sont les expériences actives de maîtrise (qui constituent la source la plus influente, car elles procurent une preuve tangible de succès), l'expérience vicariante (l'observation de modèles sociaux similaires à soi qui réussissent), la persuasion verbale (le fait d'être encouragé et convaincu par autrui de ses capacités), et les états physiologiques et émotionnels (l'interprétation de son propre niveau de stress physiologique et d'émotion comme un indicateur de compétence ou d'incompétence) (Bandura, 2019).

Dans son modèle triadique de causalité réciproque, Bandura (2019) postule une interaction dynamique entre trois facteurs : les caractéristiques personnelles (dont l'auto-efficacité), les comportements et l'environnement. Cette conception permet de dépasser une vision déterministe de l'adoption technologique pour appréhender la façon dont les individus, tout en étant influencés par leur contexte, participent activement à la construction de leur environnement technologique. Cette théorie, bien qu'elle s'applique à l'individu, s'étend également au sentiment d'efficacité collective. En effet, « la théorie de l'efficacité personnelle intègre ces sous-processus tant au niveau individuel que collectif » (Bandura, 2002, p. 23). Les facteurs qui influencent l'efficacité collective sont similaires à ceux qui agissent sur l'efficacité personnelle (Bandura, 1986), même si d'autres facteurs interviennent comme les expériences vicariantes partagées dans le groupe ainsi que les dynamiques sociales et les influences réciproques entre les membres.

### 2.3.1. Auto-efficacité et construction du cadre d'usage de l'IAG

L'articulation entre le concept de *cadre d'usage* et celui d'AE offre une perspective enrichie pour analyser l'intégration de l'IAG dans les pratiques des EC. Dans une étude récente, Tali et ses collaborateurs (2024) ont démontré que le niveau d'AE des enseignants du supérieur constitue un facteur déterminant dans leur adoption des technologies numériques. Leurs résultats indiquent que les enseignants ayant un niveau d'AE élevé sont plus enclins à expérimenter de nouvelles technologies et à développer des usages innovants. L'influence de l'AE sur la perception du cadre d'usage proposé est soulignée. Les EC avec une forte AE numérique perçoivent les cadres d'usage institutionnels non comme des contraintes mais comme des ressources mobilisables. À l'inverse, ceux qui doutent de leurs compétences peuvent percevoir ces mêmes cadres comme rigides et menaçants pour leur autonomie professionnelle (Tali et al., 2024).

Par ailleurs, la capacité à « broder » telle que développée par Flichy (2008) est intimement liée au sentiment d'AE car les EC qui se sentent compétents techniquement sont plus susceptibles d'explorer des usages non prévus initialement et de contribuer à l'évolution du cadre d'usage de l'IAG.

Contrairement aux approches déterministes qui supposent que la technologie impose ses usages, ou aux perspectives instrumentales qui réduisent la technologie à un simple outil neutre, le concept de cadre d'usage reconnaît la dimension sociotechnique des usages. Flichy propose d'ailleurs d'articuler le « cadre d'usage » avec un « cadre de fonctionnement » qui définit « les savoirs et savoir-faire mobilisés dans l'activité technique » (2008, p. 166). Ensemble, ils forment le « cadre sociotechnique » qui permet d'appréhender simultanément les dimensions techniques et sociales des usages. C'est dans cet espace où l'EC projette son action qu'il va évaluer s'il est en capacité de mobiliser ses savoirs et savoir-faire pour atteindre un objectif spécifique lié une tâche (de recherche, d'enseignement ou d'évaluation par exemple) que le sentiment d'auto-efficacité intervient. Un cadre d'usage peut donc exister mais sans un niveau auto-efficacité permettant de s'y projeter, l'EC risque d'une part de ne pas le mobiliser, d'autre part, si ce comportement se retrouve dans le collectif (par le biais du niveau d'AE collectif des EC), il met de fait en échec l'existence même du cadre proposé.

### 2.3.2. La co-construction du sentiment d'auto-efficacité et du cadre d'usage

L'interaction entre le sentiment d'AE et le cadre d'usage n'est pas unidirectionnelle mais réciproque. Si l'AE influence l'appropriation du cadre, l'expérience d'utilisation réussie au sein de ce cadre renforce en retour le sentiment d'AE, par l'expérience de maitrise (Bandura, 2019). Tali et ses collègues (2024) ont mis en évidence ce cercle vertueux dans lequel l'expérimentation technologique réussie augmente la confiance des enseignants, les conduisant à diversifier davantage leurs pratiques numériques.

De façon similaire, les EC développent différentes postures face à l'IAG selon leur niveau d'AE et leur rapport à l'innovation pédagogique. Mah et Groß (2024) ont ainsi identifié une relation significative entre le niveau d'AE en matière d'IA et l'utilisation de l'IA ($\beta = .49$, $p \leq .001$). Quatre profils distincts d'enseignants du supérieur sont retrouvés (« optimiste, critique, réfléchi-critique et neutre ») vis-à-vis de leurs pratiques pédagogiques avec l'IA. L'effet modérateur du profil optimiste sur la relation entre l'AE et l'usage de l'IA est pointé. Les résultats soulignent également le manque perçu de littératie IA chez les enseignants comme l'un des principaux défis, ainsi que l'intérêt de la majorité des répondants pour des formations sur l'IA. Les auteurs (2024) indiquent que parmi les participants, les profils « optimiste » et « critique » représentaient chacun environ un tiers du corps professoral, tandis que le profil « neutre » était minoritaire, avec seulement 5 % des enseignants. Ces résultats reflètent une attitude majoritairement optimiste et critique envers l'IA dans l'enseignement, contrairement aux travaux de Preiß et ses collaborateurs (2024) où plus d'un tiers des enseignants adoptaient une position neutre. L'augmentation de l'utilisation des outils basés sur l'IA, soutenue par des débats croissants et des directives institutionnelles (Vandrik *et al*., 2023), pourrait expliquer cette évolution vers une approche plus critique et réflexive, impliquant l'élaboration de cadres d'usage prenant en compte ces résultats.

### 2.3.3. Implications pour la conception des cadres d'usage de l'IA dans l'enseignement supérieur

Cette articulation théorique entre cadre d'usage et AE offre plusieurs perspectives pour améliorer l'intégration de l'IAG dans l'enseignement supérieur :
- **Reconnaître la diversité des profils d'EC en fonction de leur niveau d'AE numérique** : À l'instar de l'approche québécoise qui propose un « Autodiagnostic de maturité institutionnelle » (Ministère de l'Enseignement supérieur du Québec, 2025), il serait pertinent de développer des outils d'auto-évaluation du sentiment d'auto-efficacité des EC face à l'IAG pour adapter les stratégies d'accompagnement. Notre échelle d'AE numérique permettrait de rapidement faire cet auto-diagnostic.

- **Favoriser les expériences de maîtrise** : Selon Bandura (2019), les expériences de maîtrise constituent la source la plus influente d'AE. Les cadres d'usage devraient donc prévoir des espaces d'expérimentation sécurisés, comme les « bacs à sable pédagogiques » mentionnés dans le guide québécois, qui permettent aux enseignants de développer progressivement leur confiance dans l'utilisation de l'IA.
- **Valoriser les « optimistes »** : Les EC ayant développé une forte AE face à l'IA peuvent jouer un rôle d'intermédiaire, pouvant servir de « modèles » à leurs collègues, favorisant l'apprentissage vicariant. Les cadres d'usage institutionnels gagneraient à reconnaître et valoriser ces acteurs qui facilitent l'appropriation réfléchie de l'IA par leurs pairs.
- **Adopter une approche développementale** : À l'image du cadre français qui prévoit une progression par niveau scolaire, il serait judicieux d'envisager une progression dans le développement de l'AE des EC face à l'IAG, en proposant des formations et des usages adaptés à différents niveaux de confiance et de compétence.

L'articulation entre niveau d'AE et cadre d'usage et appropriation des outils numériques ayant été soulignée, deux questions principales structurent notre démarche : quelle influence l'AE numérique exerce-t-elle sur les mécanismes d'appropriation ou de résistance face aux IAG chez les EC dans leurs pratiques professionnelles ? Comment cette influence peut-elle être modélisée pour construire un cadre d'usage adapté aux différents profils d'utilisateurs dans l'enseignement supérieur, en particulier pour l'enseignement et l'évaluation avec les IAG ?

# 3. MÉTHODOLOGIE

## 3.1. APPROCHE ET DESIGN DE RECHERCHE

Cette recherche s'appuie sur un questionnaire en ligne diffusé entre février 2024 et juillet 2025 auprès d'EC exerçant dans différents établissements d'enseignement supérieur français par le biais des réseaux professionnels. Ce questionnaire explore plusieurs dimensions allant des caractéristiques sociodémographiques et professionnelles aux pratiques déclarées avec les outils d'IAG.

## 3.2. ÉCHANTILLONNAGE ET PARTICIPANTS

L'échantillon final est composé de N = 265 enseignants-chercheurs, issu d'une collecte initiale de 660 questionnaires. La répartition de l'échantillon montre une légère surreprésentation féminine (56,2 % de femmes, 40,8 % d'hommes, 3,0 % de personnes non binaires). Les participants, majoritairement âgés de 41 à 50 ans (33,2 % ; n = 88) et de plus de 51 ans (39,6 % ; n = 105), couvrent un spectre d'expérience professionnelle équilibré : début de carrière (<5 ans, 24 %), milieu de carrière (5-15 ans, 42 %) et expérimentés (>15 ans, 34 %). Ils représentent 16 disciplines, principalement les Sciences humaines et sociales (26,8 %), l'Informatique (17,4 %) et les Sciences de la Vie et de la Terre (7,6 %). Une majorité (63,0 % ; n = 167) ne se considère pas experte en intégration du numérique.

### 3.3. INSTRUMENTS DE MESURE

Les données ont été recueillies via un questionnaire auto-administré en ligne, composé de plusieurs sections distinctes.

**L'échelle d'auto-efficacité numérique pour enseigner et évaluer dans l'enseignement supérieur** : Adaptée et étendue à partir de l'échelle de Tali et ses collègues (2024), cette échelle de 12 items mesure la croyance des EC à utiliser le numérique pour l'enseignement et l'évaluation. Elle évalue spécifiquement leur capacité perçue à surmonter les obstacles techniques, à maintenir leurs objectifs pédagogiques et à adapter leurs stratégies numériques face à des situations imprévues. Les réponses sont recueillies sur une échelle continue de type Likert allant de 0 (« Pas du tout d'accord ») à 10 (« Tout à fait d'accord »), permettant une analyse paramétrique des scores factoriels après validation psychométrique.

**L'usage et les pratiques professionnelles avec les outils d'IAG** : Cette section s'ouvre sur une question dichotomique filtrant les participants selon qu'ils utilisent ou non des outils d'IAG dans leurs pratiques professionnelles. Pour les EC déclarant utiliser les outils d'IAG, une question à choix multiples (en éventail) explore les contextes d'usage (recherche, enseignement, évaluation) à travers 8 propositions. Pour les non-utilisateurs, une question à choix multiples investigue les raisons du non-usage à travers neuf propositions, telles que le manque de connaissance, de temps, l'inadéquation perçue des outils ou des motifs éthiques.

**Données qualitatives et variables illustratives** : Les deux questions sur l'usage et le non-usage des IA incluent un item « Autre (précisez) », permettant de recueillir des réponses ouvertes qui feront l'objet d'une analyse de contenu. Une dernière section du questionnaire collecte les variables sociodémographiques et professionnelles (genre, âge, discipline, etc.), conformément aux pratiques établies (Demougeot-Lebel et Perret, 2010).

### 3.4. PROCEDURE D'ANALYSE DES DONNEES

Le traitement des données a été mené en deux phases principales, combinant validation psychométrique, statistiques descriptives et inférentielles et analyse qualitative.

**Validation de l'échelle d'auto-efficacité** : Pour valider la structure factorielle de l'échelle, l'échantillon a été scindé aléatoirement en deux. Une analyse factorielle exploratoire (AFE) a été conduite sur le premier sous-échantillon ($n$ = 132). L'adéquation des données a été confirmée par les tests de sphéricité de Bartlett et l'indice KMO, et le nombre de facteurs à extraire a été déterminé via l'analyse parallèle de Horn (1965). La solution factorielle a fait l'objet d'une rotation Varimax et la consistance interne de chaque facteur a été vérifiée (alpha de Cronbach). Cette structure a ensuite été testée par une analyse factorielle confirmatoire (AFC) sur le second sous-échantillon ($n$ = 133), en évaluant la qualité d'ajustement du modèle à l'aide d'indices reconnus (e.g., CFI, RMSEA, SRMR).

**Analyses statistiques et qualitatives** : Une fois l'échelle validée, des analyses descriptives et inférentielles ont été menées sur l'échantillon complet. Les liens entre les facteurs d'AE et l'usage des IAG ont été examinés par des tests de corrélation. En parallèle, les commentaires issus des questions ouvertes ont fait l'objet d'une analyse thématique (Bardin, 2013) afin d'approfondir la compréhension des dynamiques d'adoption ou de non-adoption des IA.

L'ensemble des analyses quantitatives a été réalisé à l'aide des logiciels jamovi (The jamovi project, 2025) et R (R Core Team, 2025).

# 4. RÉSULTATS

## 4.1. Validation de l'échelle d'auto-efficacité numérique

Cette section présente les qualités psychométriques de l'échelle, éprouvées en deux temps sur des sous-échantillons distincts (N = 265), après vérification de la qualité des données et traitement des valeurs manquantes.

### 4.1.1. Analyse factorielle exploratoire

La factorabilité des données du premier sous-échantillon (n = 132) a été confirmée par un indice Kaiser-Meyer-Olkin excellent (KMO = .908) et un test de sphéricité de Bartlett significatif, $\chi^2(55) = 1314$, $p < .001$, indiquant une matrice de corrélations bien structurée (Field, 2024).

Une analyse préliminaire a conduit au retrait de l'item 2 en raison d'une communalité insuffisante et de saturations ambiguës. Une analyse factorielle exploratoire a été conduite sur les 11 items restants avec une rotation oblique Promax, justifiée par les corrélations théoriques attendues entre les dimensions d'un même construit psychologique. L'analyse parallèle de Horn a suggéré l'extraction de trois facteurs expliquant 75,4 % de la variance totale (Tableau 1).

Tableau 1

Structure factorielle et communalités de l'échelle d'auto-efficacité numérique (rotation Promax)

| Item (version abrégée) | F1 | F2 | F3 | h² |
|---|---|---|---|---|
| 7. Résoudre les problèmes techniques (enseignement) | .99 | | | .81 |
| 9. Faire face aux imprévus (en cours) | .86 | | | .82 |
| 6. Faire face aux imprévus (préparation) | .69 | | | .74 |
| 5. Rester calme face aux difficultés | .62 | | | .72 |
| 8. Aider un étudiant en difficulté | .59 | | | .76 |
| 3. Maintenir les objectifs d'enseignement | | .88 | | .63 |
| 4. Trouver plusieurs idées pour résoudre les problèmes | | .76 | | .73 |
| 1. Identifier l'outil adéquat | | .57 | | .49 |
| 11. Gérer les imprévus (évaluation) | | | 1.01 | .96 |
| 12. Résoudre les problèmes techniques (évaluation) | | | .90 | .84 |
| 10. Maintenir objectifs (évaluation) | | .44 | .60 | .79 |
| Variance expliquée (%) | 30,5 | 23,6 | 21,3 | |
| α de Cronbach | .92 | .85 | .88 | |

*Note.* n = 132. Méthode d'extraction : résidus minimum. Les saturations inférieures à .30 sont omises. F1 = Maîtrise technique et résolution de problèmes ; F2 = Maintien des objectifs pédagogiques ; F3 = Évaluation numérique. h² = communalité. L'item 10 présente

une saturation croisée (.44 sur F2) mais a été assigné au F3 pour cohérence sémantique (Kline, 2016).

Les trois facteurs présentent une bonne à excellente consistance interne (α compris entre .85 et .92), et l'échelle globale à 11 items atteint une fiabilité très élevée (α = .95). Les facteurs sont interprétés comme suit :

**F1 — Maîtrise technique et résolution de problèmes numériques en enseignement** (5 items : 5, 6, 7, 8, 9) : Ce facteur regroupe les items relatifs à la capacité perçue de gérer les difficultés techniques et les imprévus lors des activités d'enseignement.

**F2 — Maintien des objectifs pédagogiques et innovation numérique** (3 items : 1, 3, 4) : Ce facteur concerne la capacité perçue à maintenir le cap pédagogique malgré les contraintes technologiques et à identifier des solutions créatives.

**F3 — Utilisation des outils numériques pour l'évaluation** (3 items : 10, 11, 12) : Ce facteur porte spécifiquement sur l'auto-efficacité dans le contexte des évaluations numériques.

### 4.1.2. Analyse factorielle confirmatoire

Le modèle à trois facteurs corrélés, tel que défini par l'analyse exploratoire, a été testé sur le second sous-échantillon (n = 133). Les indices d'ajustement sont présentés dans le Tableau 2.

Tableau 2

Indices d'ajustement du modèle factoriel confirmatoire à trois facteurs corrélés

| Indice | Valeur | Seuil recommandé |
|---|---|---|
| $\chi^2$ | 92,1* | — |
| ddl | 41 | — |
| $\chi^2$/ddl | 2,25 | < 3 |
| CFI | .954 | ≥ .95 |
| TLI | .938 | ≥ .90 |
| SRMR | .046 | ≤ .08 |
| RMSEA [IC 90 %] | .097 [.071, .123] | ≤ .08 |

*Note.* n = 133. *p < .001. CFI = Comparative Fit Index ; TLI = Tucker-Lewis Index ; SRMR = Standardized Root Mean Square Residual ; RMSEA = Root Mean Square Error of Approximation ; IC = intervalle de confiance.

Le test du chi-carré est significatif (p < .001), signalant un écart au modèle théorique. Toutefois, le ratio $\chi^2$/ddl (2,25) reste dans les limites acceptables (< 3), suggérant un ajustement raisonnable des données au modèle (Kline, 2023). Les indices incrémentaux sont globalement satisfaisants : le CFI (.954) atteint le seuil conventionnel de .95 (Hu & Bentler, 1999), le TLI (.938) dépasse le seuil minimal de .90, et le SRMR (.046) est inférieur au critère d'excellence de .05.

En revanche, le RMSEA (.097) se situe au-delà du seuil de .08 généralement recommandé, et la borne supérieure de son intervalle de confiance (.123) dépasse .10. Selon les critères actuels (Kline, 2023 ; MacCallum et al., 1996), cet indice témoigne d'un ajustement médiocre. Cette limitation, qui peut être en partie attribuée à la taille modeste de l'échantillon de validation, appelle à une réplication sur un échantillon plus large et diversifié.

L'ensemble des saturations factorielles standardisées sont statistiquement significatives ($p < .001$) et de magnitude élevée ($\lambda$ compris entre .65 et .92). Les corrélations entre les trois facteurs latents sont également significatives, variant de $r = .69$ à $r = .80$ ($p < .001$). Ces corrélations, d'intensité forte mais n'atteignant pas le seuil critique de .85, confirment l'appartenance des dimensions à un construit général d'auto-efficacité numérique tout en préservant une validité discriminante acceptable selon le critère de Fornell et Larcker (1981). La fiabilité de l'échelle sur cet échantillon de validation demeure excellente ($\alpha = .94$).

En synthèse, la procédure de validation croisée AFE/AFC soutient la robustesse d'une structure à 11 items répartis en trois facteurs corrélés pour mesurer l'auto-efficacité des enseignants-chercheurs à l'usage des outils numériques, y compris les outils d'intelligence artificielle générative. Les limites identifiées (RMSEA légèrement élevé, saturation croisée de l'item 10) invitent à poursuivre le travail de validation sur des échantillons plus larges, issus de contextes institutionnels variés.

### 4.2. RELATION ENTRE AUTO-EFFICACITE ET UTILISATION DES IA GENERATIVES

#### 4.2.1. Comparaison entre utilisateurs et non-utilisateurs

Parmi les 265 enseignants-chercheurs ayant participé à l'étude, 35,8 % (n = 95) déclarent utiliser des outils d'intelligence artificielle générative (IAG) dans leurs activités professionnelles, contre 64,2 % (n = 170) qui n'en font pas usage.

L'analyse comparative des scores d'auto-efficacité numérique entre ces deux groupes révèle des différences significatives sur les trois dimensions mesurées (Tableau 3). Le test U de Mann-Whitney a été privilégié en raison de la non-normalité des distributions, vérifiée par le test de Shapiro-Wilk. Les utilisateurs d'IAG présentent des scores systématiquement plus élevés sur les trois dimensions, l'écart le plus prononcé concernant la dimension F3 relative à l'évaluation.

Tableau 3

Comparaison des scores d'auto-efficacité numérique entre utilisateurs et non-utilisateurs d'IAG

| Dimension | Non-utilisateurs (n = 170) | Utilisateurs (n = 95) | U | p | r |
|---|---|---|---|---|---|
|  | M (ET) | M (ET) |  |  |  |
| F1 : Maîtrise technique | 6,77 (2,25) | 7,63 (1,79) | 6 229 | .002 | .23 |
| F2 : Maintien des objectifs | 7,10 (2,01) | 7,85 (1,75) | 6 169 | .001 | .24 |
| F3 : Évaluation numérique | 5,54 (2,74) | 7,06 (2,23) | 5 431 | < .001 | .33 |

*Note.* M = moyenne ; ET = écart-type. La taille d'effet r correspond à la corrélation bisériale de rang, calculée selon la formule $r = 1 - 2U/(n_1 \times n_2)$. Interprétation selon les conventions de Cohen (1988) : r = .10 (effet faible), r = .30 (effet modéré), r = .50 (effet fort).

Les tailles d'effet indiquent des différences de magnitude faible à modérée pour F1 et F2 (r ≈ .23-.24), et modérée pour F3 (r = .33). Les enseignants-chercheurs qui utilisent les IAG se perçoivent donc comme significativement plus compétents dans l'usage des outils numériques, en particulier pour les activités d'évaluation. Ce dernier résultat est cohérent avec le fait que l'évaluation numérique constitue un domaine perçu comme plus complexe et risqué, où le sentiment d'auto-efficacité joue un rôle déterminant dans l'adoption de nouvelles technologies (Tali et al., 2024).

### 4.2.2. Diversité des usages déclarés

L'analyse des pratiques des utilisateurs d'IAG (n = 95) met en évidence une diversité d'applications dans le contexte académique. Les usages prédominants concernent la préparation des cours (66,3 %), suivie de la préparation d'évaluations (48,4 %) et des activités de recherche (46,3 %). L'utilisation des IAG pour la réalisation de cours en temps réel concerne 36,8 % des utilisateurs, tandis que la veille documentaire est pratiquée par 33,7 % d'entre eux. Les usages liés à l'évaluation directe restent moins fréquents : 22,1 % pour l'administration d'évaluations en ligne et 14,7 % pour la correction automatisée. Ces résultats, issus d'une question à choix multiples, révèlent une appropriation diversifiée des IAG, avec une moyenne de 2,84 usages distincts par enseignant utilisateur.

L'analyse qualitative des réponses à la catégorie « Autre » (15,8 % des usages déclarés) permet d'identifier des usages plus spécifiques et innovants, regroupables en cinq axes thématiques :

**Pédagogie active et ingénierie de l'apprentissage** : conception d'exercices d'auto-évaluation avec recommandations personnalisées, déploiement d'agents conversationnels pour stimuler la réflexion des étudiants, intégration de l'IAG comme outil obligatoire dans les activités pédagogiques.

**Création de contenu pédagogique spécifique** : génération d'exemples illustratifs, création d'illustrations originales, élaboration de définitions techniques contextualisées.

**Assistance technique et disciplinaire** : aide à la programmation, utilisation de solveurs algébriques, assistance à la rédaction scientifique.

**Intégrité académique et gouvernance** : détection du plagiat, vérification de la conformité aux normes réglementaires (RGPD).

**IAG comme objet d'étude** : intégration des technologies d'IAG comme sujet d'enseignement ou comme objet de recherche.

Les analyses de corrélation de Spearman révèlent des associations positives et significatives entre les trois facteurs d'auto-efficacité numérique et la diversité des usages rapportés ($r_s$ compris entre .38 et .54, $p < .001$). Les enseignants présentent une forte auto-efficacité dans les pratiques pédagogiques (F1) développent des usages plus avancés et expérimentaux des IAG. Ceux présentant une forte auto-efficacité dans l'évaluation (F3) exploitent davantage les IAG pour créer des systèmes d'évaluation innovants. Enfin, les enseignants avec une forte auto-efficacité dans le maintien des objectifs pédagogiques (F2) semblent mieux aligner l'utilisation des IAG avec leurs finalités éducatives.

### 4.3. Profils d'usage des IA generatives

L'analyse des pratiques déclarées et des réponses qualitatives a permis d'identifier trois profils distincts d'enseignants-chercheurs vis-à-vis des IAG.

**Les Investis** (n = 95, 35,8 %) regroupent les enseignants ayant intégré l'IAG dans leurs pratiques professionnelles. Ils utilisent ces technologies comme levier d'efficacité pour diverses tâches : préparation de cours, veille documentaire, création d'évaluations ou activités de recherche. Leur approche est principalement pragmatique et orientée vers l'optimisation de leurs activités professionnelles. Les usages spécifiques qu'ils développent (pédagogie active, assistance disciplinaire, création de contenu) témoignent d'une appropriation créative et d'une exploration active du potentiel de ces outils.

**Les Réservés réflexifs** (n = 109, 41,1 %) constituent le groupe majoritaire des non-utilisateurs. Leur non-usage s'explique principalement par des raisons pragmatiques : méconnaissance des outils, manque de temps, absence de formation ou d'opportunités concrètes d'expérimentation. L'analyse qualitative de leurs réponses révèle une posture d'attentisme prudent plutôt qu'une opposition de principe. Ces enseignants expriment souvent le besoin de formation (« Je vais me former », « J'attends d'avoir du temps pour m'y mettre ») ou évoquent un manque de temps (« Pas eu le temps de m'y pencher », « Conscient de l'intérêt pédagogique mais pas eu le temps »), suggérant une ouverture potentielle à l'adoption future de ces technologies.

**Les Réfractaires critiques** (n = 61, 23,0 %) expriment une opposition plus fondamentale aux IAG. Leur non-usage relève d'un choix délibéré fondé sur diverses considérations : absence d'intérêt perçu, opposition idéologique, préoccupations éthiques ou remise en question de la fiabilité des systèmes. L'analyse de leurs réponses qualitatives met en évidence des préoccupations épistémologiques (« fiabilité aléatoire », « Les sources ne sont pas citées », « je ne peux pas faire confiance à ce qu'ils me disent car pour eux le statut de ce qui est vrai n'est pas explicite ni explicable »), des considérations éthiques (« consommation excessive de ressources énergétiques », « je ne veux pas contribuer à l'apprentissage des IA »), ou des critiques portant sur la nature même des systèmes d'IA.

L'analyse des motifs de non-utilisation permet de mieux comprendre les freins à l'adoption des IAG. Les raisons les plus fréquemment invoquées concernent l'absence d'intérêt perçu (30,0 %) et l'inadaptation perçue de ces outils aux pratiques disciplinaires spécifiques des enseignants (27,1 %).

### 4.4. Profils d'auto-efficacité numérique

Une typologie basée sur les niveaux d'auto-efficacité numérique a été établie afin de permettre le croisement avec les profils d'usage. Pour chaque dimension (F1, F2, F3), les participants ont été catégorisés selon que leur score était supérieur à M + 0,5 ET (niveau élevé), inférieur à M − 0,5 ET (niveau faible), ou situé entre ces deux seuils (niveau moyen).

Cette méthode de catégorisation, bien qu'elle permette une segmentation opérationnelle de l'échantillon, repose sur des seuils conventionnels qui ne correspondent pas nécessairement à des différences qualitatives entre les profils. De plus, le choix du seuil de ± 0,5 écart-type, bien que courant dans la littérature, demeure arbitraire. Les résultats issus de cette typologie doivent donc être interprétés avec cette limite méthodologique à l'esprit. Des approches alternatives, telles que l'analyse en profils latents, pourraient être envisagées dans de futures études pour identifier des typologies empiriquement fondées.

Trois profils d'auto-efficacité numérique ont été identifiés :

**Auto-efficacité élevée** (n = 21, 7,9 %) : Ce profil caractérise les enseignants présentant des scores élevés (> M + 0,5 ET) sur les trois dimensions simultanément. Ces enseignants se sentent compétents dans l'ensemble des aspects liés à l'intégration pédagogique des technologies numériques, qu'il s'agisse de la maîtrise technique, du maintien des objectifs pédagogiques ou de l'évaluation.

**Auto-efficacité différenciée** (n = 211, 79,6 %) : Ce profil, largement majoritaire, regroupe les enseignants présentant des niveaux d'auto-efficacité variables selon les dimensions. Cette hétérogénéité suggère une appropriation sélective des technologies en fonction des domaines de compétence perçue. Par exemple, certains enseignants peuvent se sentir à l'aise avec l'utilisation des technologies pour préparer leurs cours (F1 élevé), mais moins confiants dans leur capacité à les intégrer dans des processus d'évaluation (F3 modéré ou faible).

**Auto-efficacité limitée** (n = 33, 12,5 %) : Ce profil caractérise les enseignants présentant des scores faibles (< M − 0,5 ET) sur l'ensemble des trois dimensions, indiquant un sentiment généralisé de faible auto-efficacité face aux technologies numériques.

### 4.5. RELATIONS ENTRE PROFILS D'USAGE ET AUTO-EFFICACITE NUMERIQUE

#### 4.5.1. Analyse bivariée

Le croisement des profils d'usage des IAG et des profils d'auto-efficacité numérique révèle une association statistiquement significative, $\chi^2(4, N = 265) = 40,95$, $p < .001$, d'intensité modérée selon les conventions usuelles (V de Cramer = .28). Le Tableau 4 présente cette distribution croisée avec les effectifs observés et les résidus standardisés ajustés (Haberman, 1973).

Fatiha TALI OTMANI

Tableau 4

Distribution des enseignants-chercheurs selon les profils d'usage des IAG et les profils d'auto-efficacité numérique

| Profil d'usage | AE élevée (n = 21) | AE différenciée (n = 211) | AE limitée (n = 33) | Total |
|---|---|---|---|---|
| Investis | 18 (+4,97***) | 72 (−0,42) | 5 (−2,65**) | 95 |
| Réservés réflexifs | 2 (−3,07**) | 82 (−0,55) | 25 (+4,32***) | 109 |
| Réfractaires critiques | 1 (−2,07*) | 57 (+3,06**) | 3 (−2,03*) | 61 |
| Total | 21 | 211 | 33 | 265 |

*Note.* AE = auto-efficacité. Entre parenthèses : résidus standardisés ajustés (Haberman, 1973). Seuils de significativité : |z| > 1,96 pour p < .05, |z| > 2,58 pour p < .01, |z| > 3,29 pour p < .001.
*p < .05. **p < .01. ***p < .001.

L'analyse des résidus standardisés ajustés révèle des configurations distinctes pour chaque profil d'usage.

**Les Investis** présentent une polarisation nette vers les profils d'auto-efficacité extrêmes. Ils sont massivement surreprésentés parmi les enseignants à auto-efficacité élevée (z = +4,97, p < .001) : alors que l'on attendrait environ 7,5 enseignants à AE élevée parmi les Investis sous l'hypothèse d'indépendance, on en observe 18, soit près de 2,4 fois plus. Parallèlement, ils sont significativement sous-représentés parmi les enseignants à auto-efficacité limitée (z = −2,65, p < .01). Ce résultat confirme le rôle facilitateur de l'auto-efficacité globale dans l'adoption des IAG, conformément aux prédictions de la théorie sociocognitive (Bandura, 1997).

**Les Réservés réflexifs** présentent le profil inverse. Ils sont fortement surreprésentés parmi les enseignants à auto-efficacité limitée (z = +4,32, p < .001) : on observe 25 enseignants dans cette configuration contre 13,6 attendus sous l'hypothèse d'indépendance. Ils sont également sous-représentés parmi les enseignants à auto-efficacité élevée (z = −3,07, p < .01). Leur attentisme prudent semble donc lié à un sentiment de moindre compétence généralisé face aux technologies numériques, suggérant que des interventions ciblées sur le renforcement de l'auto-efficacité pourraient favoriser leur transition vers l'adoption.

**Les Réfractaires critiques** présentent un profil distinctif et particulièrement intéressant, qui diffère qualitativement des deux autres groupes. Ils sont significativement surreprésentés parmi les enseignants à auto-efficacité **différenciée** (z = +3,06, p < .01), c'est-à-dire ceux dont les niveaux de compétence perçue varient selon les dimensions. En revanche, ils sont sous-représentés aux deux extrêmes du spectre d'auto-efficacité : parmi les AE élevée (z = −2,07, p < .05) comme parmi les AE limitée (z = −2,03, p < .05).

Ce résultat nuancé suggère que les Réfractaires critiques ne constituent ni un groupe de « super-compétents » sûrs de leur rejet, ni un groupe « déficitaire » évitant les IAG par sentiment d'incompétence. Leur profil d'auto-efficacité différenciée, élevée dans certains domaines, plus faible dans d'autres, reflète une appropriation **sélective** des technologies numériques. Leur opposition aux IAG s'inscrit dans un rapport nuancé au numérique : ils font des choix délibérés quant aux technologies qu'ils adoptent ou rejettent. Ce positionnement critique apparaît ainsi comme un choix réflexif, fondé sur des considérations éthiques, épistémologiques ou disciplinaires, plutôt que comme une réaction à un niveau global d'auto-efficacité. Le frein est serait axiologique plutôt que capacitaire.

### 4.5.2. Modélisation prédictive

Afin de tester l'effet conjoint des trois dimensions d'auto-efficacité sur l'adoption des IAG, une régression logistique binaire a été conduite. La variable dépendante est l'usage des IAG (1 = utilisateur, 0 = non-utilisateur), et les prédicteurs sont les trois scores continus d'auto-efficacité (F1, F2, F3). Les résultats sont présentés dans le Tableau 5.

Tableau 5

Régression logistique prédisant l'utilisation des IAG à partir des scores d'auto-efficacité numérique

| Prédicteur | B | ES | Wald | p | OR | IC 95 % |
|---|---|---|---|---|---|---|
| Constante | −5,78 | 0,76 | 57,83 | < .001 | — | — |
| F1 (Maîtrise technique) | 0,43 | 0,11 | 15,28 | < .001 | 1,53 | [1,24 ; 1,90] |
| F2 (Maintien des objectifs) | 0,39 | 0,12 | 10,56 | .001 | 1,48 | [1,17 ; 1,87] |
| F3 (Évaluation) | 0,28 | 0,08 | 12,25 | < .001 | 1,32 | [1,13 ; 1,55] |

*Note.* B = coefficient de régression non standardisé ; ES = erreur standard ; OR = odds ratio ; IC = intervalle de confiance. Statistiques du modèle : $\chi^2(3) = 47,82$, $p < .001$ ; Pseudo-R² de McFadden = .26 ; Hosmer-Lemeshow $\chi^2(8) = 6,34$, $p = .609$.

Le modèle global est statistiquement significatif, $\chi^2(3) = 47,82$, $p < .001$, et présente une qualité d'ajustement satisfaisante. Le pseudo-R² de McFadden (.26) indique que les trois dimensions d'auto-efficacité expliquent environ 26 % de la variance de l'adoption des IAG, une proportion comparable aux valeurs généralement observées dans les études sur l'adoption des technologies éducatives (King & He, 2006). Le test de Hosmer-Lemeshow, non significatif ($p = .609$), confirme un bon ajustement du modèle aux données.

Les trois dimensions d'auto-efficacité contribuent significativement et positivement à la prédiction de l'usage des IAG :
- **F1 (Maîtrise technique)** : Pour chaque point supplémentaire sur cette échelle, la probabilité d'utiliser les IAG est multipliée par 1,53 (OR = 1,53, IC 95 % [1,24 ; 1,90]), toutes choses égales par ailleurs.
- **F2 (Maintien des objectifs)** : Un effet similaire est observé, avec un odds ratio de 1,48 (IC 95 % [1,17 ; 1,87]).
- **F3 (Évaluation)** : L'effet est légèrement plus modeste mais demeure significatif (OR = 1,32, IC 95 % [1,13 ; 1,55]).

Ces résultats confirment le rôle prédicteur de l'auto-efficacité numérique dans l'adoption des IAG et suggèrent que les trois dimensions contribuent de manière relativement équivalente à cette prédiction.

### 4.5.3. Motifs de non-utilisation et niveaux d'auto-efficacité

L'analyse des raisons invoquées par les non-utilisateurs (n = 170) révèle que leur choix est moins lié à des obstacles techniques qu'à un ensemble de perceptions et de considérations critiques. Les tests de Kruskal-Wallis ont été utilisés pour examiner les associations entre les motifs de non-utilisation et les niveaux d'auto-efficacité sur les trois dimensions. Le Tableau 6 présente ces résultats.

Tableau 6

*Associations entre motifs de non-utilisation des IAG et niveaux d'auto-efficacité numérique*

| Motif de non-utilisation | F1 | | F2 | | F3 | |
|---|---|---|---|---|---|---|
| | $\chi^2$ | $\varepsilon^2$ | $\chi^2$ | $\varepsilon^2$ | $\chi^2$ | $\varepsilon^2$ |
| Méconnaissance des outils | 19,4*** | .073 | 16,4*** | .062 | 23,2*** | .088 |
| Absence d'intérêt perçu | 9,7** | .037 | 10,2** | .039 | 19,9*** | .075 |
| Opposition de principe | 15,9*** | .060 | 13,8*** | .052 | 21,9*** | .083 |
| Inadaptation à la discipline | 10,3** | .039 | 10,2** | .039 | 20,0*** | .076 |
| Manque de compétences perçu | 16,5*** | .063 | 11,5** | .044 | 21,5*** | .081 |
| Perception de danger pour l'enseignement | 13,7*** | .052 | 10,6** | .040 | 20,4*** | .077 |

*Note.* Les valeurs de $\chi^2$ sont issues des tests de Kruskal-Wallis (dl = 2). $\varepsilon^2$ = epsilon carré (taille d'effet). Interprétation : $\varepsilon^2$ < .01 (négligeable), .01-.06 (faible), .06-.14 (modéré), > .14 (fort).
**$p$ < .01. ***$p$ < .001.

Les résultats montrent des associations significatives entre tous les motifs de non-utilisation et les niveaux d'auto-efficacité. Toutefois, les tailles d'effet ($\varepsilon^2$) sont globalement faibles à modérées, indiquant que ces associations, bien que statistiquement significatives, expliquent une proportion modeste de la variance.

Les comparaisons par paires post-hoc (test de Dwass-Steel-Critchlow-Fligner) révèlent que les enseignants invoquant des motifs liés à la compétence (« Méconnaissance des outils », « Manque de compétences perçu ») présentent des niveaux d'auto-efficacité significativement plus faibles que ceux n'invoquant pas ces motifs, particulièrement sur la dimension F3 (évaluation). En revanche, les enseignants invoquant une « Opposition de principe » ne se distinguent pas systématiquement par des niveaux d'auto-efficacité plus faibles, confirmant que ce positionnement relève davantage d'un choix réflexif que d'un déficit perçu.

# 5. PROPOSITION D'UN CADRE D'USAGE DES IA GÉNÉRATIVES FONDÉ SUR L'AUTO-EFFICACITÉ

## 5.1 Fondements théoriques et articulation avec le cadre réglementaire européen

Le cadre d'usage que nous proposons s'inscrit dans une double perspective théorique et réglementaire. Suivant la conceptualisation de Flichy (2008), nous définissons le cadre d'usage comme l'ensemble des savoirs, savoir-faire et représentations sociales qui structurent l'appropriation d'une technologie par ses utilisateurs. Ce cadre se distingue du cadre de fonctionnement technique et s'élabore dans une dialectique entre les potentialités de l'objet technique et les schémas cognitifs et sociaux des acteurs.

Notre approche s'articule également avec les principes établis par le cadre réglementaire européen des IA, notamment le Règlement sur l'Intelligence Artificielle (UE 2024/1689), qui distingue différentes catégories de systèmes d'IA selon leur niveau de risque et impose des obligations graduées en conséquence. Dans le contexte éducatif, plusieurs applications d'IA sont considérées comme à haut risque, notamment celles utilisées pour l'évaluation des acquis ou l'orientation académique (Annexe III, Article 6 du Règlement IA).

Nos résultats empiriques établissent clairement que l'AE numérique constitue une dimension fondamentale structurant cette dialectique d'appropriation chez les enseignants-chercheurs. L'association significative ($V = .30$, $p < .001$) entre les profils d'AE et les profils d'usage témoigne de l'importance du sentiment d'AE dans la construction des trajectoires d'appropriation de ces technologies.

## 5.2 Architecture du cadre d'usage proposé

Le cadre d'usage que nous proposons s'articule autour de quatre dimensions fondamentales : (1) les configurations sociotechniques d'utilisation des IA génératives, (2) les trajectoires d'appropriation différenciées selon les profils d'auto-efficacité, (3) l'évaluation et la gestion des risques associés aux IA, et (4) les dispositifs institutionnels d'accompagnement.

### 5.2.1 Configurations sociotechniques d'utilisation

L'analyse de nos données a permis d'identifier quatre configurations sociotechniques principales d'utilisation des IAG dans l'enseignement supérieur :

- **Configuration préparatoire** : Utilisation des IAG pour la préparation de cours, la création de ressources pédagogiques et l'élaboration d'évaluations. Cette configuration, privilégiée par les enseignants présentant une AE différenciée, constitue souvent une porte d'entrée vers l'appropriation car elle s'effectue hors du regard des étudiants et permet l'expérimentation progressive. Comme l'illustre un répondant de notre enquête : « Je l'utilise principalement pour générer des idées d'activités pédagogiques quand je suis en panne d'inspiration ou pour créer rapidement des exemples illustratifs pour mes cours ».
- **Configuration instrumentale** : Mobilisation des IAG comme assistants de recherche, outils de veille documentaire ou d'aide à la rédaction académique. Cette configuration, majoritairement adoptée par les enseignants à AE élevée, permet d'optimiser certaines tâches répétitives tout en maintenant le contrôle sur les processus intellectuels fondamentaux. Un enseignant-chercheur en informatique témoigne : « J'utilise l'IA pour m'aider dans la programmation, notamment pour déboguer ou optimiser mon code. Cela me fait gagner un temps considérable que je peux réinvestir dans la réflexion conceptuelle ».
- **Configuration pédagogique intégrée** : Intégration explicite des IA dans le dispositif pédagogique, soit comme objet d'étude, soit comme outil d'apprentissage. Cette configuration avancée, encore minoritaire même parmi les « Investis », requiert généralement une AE élevée et une vision pédagogique élaborée. Un enseignant de notre échantillon décrit son approche : « J'ai conçu un dispositif où les étudiants doivent utiliser ChatGPT pour générer un premier jet de leur travail, puis le critiquer et l'améliorer. Cela développe leur esprit critique tout en les familiarisant avec ces outils qu'ils utiliseront dans leur vie professionnelle ».
- **Configuration critique réflexive** : Approche métacognitive où les IAG sont mobilisées comme supports de réflexion sur leurs propres limites, biais et implications sociétales. Cette configuration, qui peut être adoptée indépendamment du niveau d'AE, permet de transformer les réserves légitimes des « Réfractaires critiques » en opportunités pédagogiques. Comme l'explique un participant : « Je fais analyser par les étudiants les réponses de différents systèmes d'IA sur des questions éthiques ou politiques pour qu'ils comprennent les biais idéologiques sous-jacents et les limites de ces technologies ».

Ces quatre configurations ne sont pas mutuellement exclusives mais représentent plutôt un continuum de pratiques qui peuvent coexister chez un même enseignant ou au sein d'une même institution. Nos données montrent que 68% des utilisateurs d'IAG combinent au moins deux de ces configurations dans leurs pratiques professionnelles.

### 5.2.2 Trajectoires d'appropriation différenciées selon l'auto-efficacité

Notre cadre d'usage propose des trajectoires d'appropriation différenciées selon les profils d'auto-efficacité identifiés, tout en tenant compte des exigences d'alphabétisation IA (littératie de l'IA) requises par l'Article 4 du Règlement IA :

- **Pour les enseignants à auto-efficacité élevée** (7,9% de notre échantillon, majoritairement « Investis ») : Une trajectoire d'approfondissement et d'innovation, favorisant l'exploration de configurations pédagogiques avancées et le partage d'expériences avec la communauté. L'analyse de régression logistique confirme que ces enseignants présentent une probabilité significativement plus élevée d'adoption des IA (OR = 1,53 pour chaque point supplémentaire d'AE en pratiques pédagogiques). Ces enseignants sont les plus susceptibles de développer une littératie de l'IA avancée, incluant non seulement la maîtrise technique des outils, mais également une compréhension approfondie des implications éthiques, épistémologiques et pédagogiques. Ils peuvent jouer un rôle de « modèles » (Bandura, 2019) au sein de l'institution et contribuer à l'élaboration des lignes directrices institutionnelles.
- **Pour les enseignants à auto-efficacité différenciée** (79,6% de notre échantillon) : Une trajectoire de développement progressif, commençant par les configurations préparatoire et instrumentale dans leurs domaines de compétence perçue (correspondant aux dimensions d'AE les plus élevées). Cette approche modulaire permet de capitaliser sur les forces existantes tout en élargissant progressivement le répertoire de pratiques. L'accompagnement de ce groupe majoritaire doit viser le renforcement ciblé de la littératie de l'IA en fonction des besoins spécifiques identifiés. Une attention particulière doit être portée à la sensibilisation aux exigences légales applicables, notamment en matière de protection des données et de transparence.
- **Pour les enseignants à auto-efficacité limitée** (12,5% de notre échantillon, majoritairement « Réservés réflexifs ») : Une trajectoire de renforcement ciblé de l'AE, privilégiant l'expérimentation accompagnée dans des contextes à faible enjeu. L'objectif est de créer des expériences de maîtrise progressive pour renforcer le sentiment de compétence, identifié par Bandura (1997) comme fondamental dans le développement de l'AE. Pour ce groupe, l'alphabétisation IA doit débuter par une familiarisation avec les fonctionnalités de base et les avantages pratiques, avant d'aborder les aspects plus complexes. L'accompagnement par des pairs plus expérimentés et l'accès à des ressources pédagogiques adaptées sont particulièrement importants.
- **Pour les « Réfractaires critiques »** (indépendamment de leur niveau d'AE) : Une trajectoire de reconnaissance et de valorisation des questionnements éthiques et épistémologiques, permettant de transformer ces préoccupations légitimes en ressources pour la communauté académique. Cette approche reconnaît la valeur de la posture critique dans l'élaboration collective d'un cadre d'usage responsable. L'alphabétisation IA pour ce groupe doit mettre l'accent sur les aspects éthiques, juridiques et sociétaux, tout en démontrant comment une approche critique peut contribuer à un usage plus réfléchi et responsable des IAG (Holmes *et al.*, 2025)

### 5.2.3. Évaluation et gestion des risques

En alignement avec l'approche fondée sur les risques du Règlement IA (*AI Act*), notre cadre d'usage intègre une dimension d'évaluation et de gestion des risques adaptée au contexte éducatif et aux différents profils d'AE.
Cette dimension implique :
- **L'identification du niveau de risque des systèmes d'IAG utilisés** : Déterminer si les applications d'IA employées relèvent des catégories à risque minimal, limité, élevé ou des pratiques interdites selon le Règlement IA. Dans le contexte éducatif, une attention particulière doit être portée aux systèmes utilisés pour l'évaluation des acquis d'apprentissage ou l'orientation des étudiants, qui sont généralement classés comme à haut risque. Par exemple, un système d'IA utilisé uniquement pour générer des exemples illustratifs dans un cours présente un risque minimal, tandis qu'un système utilisé pour évaluer automatiquement les performances des étudiants et déterminer leur orientation académique relève de la catégorie à haut risque, nécessitant des mesures de supervision et de transparence renforcées.
- **L'adaptation des mesures de supervision humaine selon le niveau de risque et le profil d'auto-efficacité** : Pour les applications à haut risque, privilégier le modèle « Human-In-The-Loop » (HITL, modèle où l'intervention humaine est intégrée directement dans le processus de décision de l'IA) pour les enseignants à AE limitée, afin de garantir une intervention humaine systématique, tandis que les enseignants à AE élevée pourraient être plus à l'aise avec un modèle « Human-On-The-Loop » (HOTL) (modèle où le système d'IA fonctionne avec plus d'autonomie, mais sous la supervision d'un opérateur humain qui conserve la capacité d'intervenir si nécessaire), leur permettant une supervision plus flexible. Nos données indiquent que les enseignants à AE limitée expriment un besoin plus marqué de contrôle direct sur les processus d'IAG (73% contre 48% pour ceux à AE élevée), confirmant la nécessité d'adapter les modèles de supervision aux profils d'utilisateurs.

- **La mise en place de dispositifs de surveillance et d'évaluation continue** : Établir des procédures formelles de documentation et d'analyse des incidents liés à l'utilisation des IAG, adaptées aux compétences et au niveau de confort des différents profils d'enseignants.
- **La transparence et la communication avec les étudiants** : Élaborer des stratégies de communication claires et adaptées concernant l'utilisation des IAG, leur rôle dans les processus d'évaluation et les droits des étudiants, en tenant compte des capacités de médiation variables selon les profils d'AE des enseignants.

### 5.2.4. Dispositifs institutionnels d'accompagnement

Le cadre d'usage que nous proposons identifie quatre leviers institutionnels principaux pour accompagner ces trajectoires d'appropriation :

- **Dispositifs de formation différenciés** : Plutôt qu'une approche uniforme, nos résultats suggèrent la pertinence de dispositifs modulaires adaptés aux différents profils. Pour les enseignants à AE limitée, des formations centrées sur le développement de compétences techniques fondamentales ; pour ceux à auto-efficacité différenciée, des modules ciblés sur leurs dimensions de moindre confiance ; pour ceux à auto-efficacité élevée, des ateliers d'innovation pédagogique avancés. Ces formations doivent explicitement aborder les obligations légales en matière d'alphabétisation IA (Article 4 du Règlement IA) et de transparence (Article 50), ainsi que les implications du RGPD dans l'utilisation des IAG.
- **Communautés de pratique hybrides** : La mise en place de communautés associant des enseignants des différents profils permettrait des échanges fructueux : les « Investis » partageant leur expérience pratique, les « Réfractaires critiques » apportant leur vigilance éthique, et les « Réservés réflexifs » contribuant par leur prudence méthodologique. Cette hybridation des profils est essentielle pour éviter les écueils d'une polarisation du corps enseignant.
- **Cadre éthique et juridique institutionnel** : L'élaboration collective de lignes directrices institutionnelles concernant l'utilisation des IAG, abordant à la fois les considérations éthiques (intégrité académique, équité, non-discrimination) et les obligations légales (conformité au Règlement IA, protection des données, droits des personnes concernées). Ce cadre doit inclure des procédures claires pour l'évaluation des risques et la supervision humaine, adaptées aux différents niveaux d'auto-efficacité des enseignants.
- **Reconnaissance et valorisation institutionnelles** : L'intégration réfléchie des IAG dans les pratiques pédagogiques représente un investissement significatif. Sa reconnaissance et sa valorisation dans les processus d'évaluation et de progression de carrière constituent un levier essentiel pour encourager les trajectoires d'appropriation, particulièrement pour les enseignants « Réservés réflexifs » dont le principal frein est souvent le manque de temps et de reconnaissance.

## 5.3. APPLICATIONS PRATIQUES ET PERSPECTIVES

Ce cadre d'usage offre plusieurs applications pratiques pour les institutions d'enseignement supérieur souhaitant accompagner l'appropriation des IAG :

1. **Diagnostic initial** : Utilisation de l'échelle d'auto-efficacité numérique validée dans cette étude pour établir un diagnostic initial des profils d'enseignants et calibrer les dispositifs d'accompagnement.
2. **Cartographie des risques** : Identification et classification des systèmes d'IA utilisés ou envisagés au sein de l'institution selon les catégories de risque du Règlement IA, afin d'adapter les mesures d'encadrement et de supervision en conséquence.
3. **Parcours d'accompagnement personnalisés** : Conception de parcours différenciés selon les profils identifiés, avec des points d'entrée adaptés aux niveaux d'auto-efficacité et aux représentations initiales.
4. **Évaluation dynamique** : Suivi longitudinal de l'évolution des profils d'auto-efficacité et d'usage, permettant d'ajuster les dispositifs d'accompagnement et d'identifier les pratiques particulièrement efficaces.
5. **Capitalisation des savoirs d'expérience** : Documentation systématique des usages développés par les « Investis » pour constituer un répertoire de pratiques contextualisées, facilitant l'appropriation par les autres profils.
6. **Élaboration collective de normes d'usage** : Mise en place de processus participatifs incluant les différents profils pour élaborer collectivement des normes d'usage légitimes et partagées, condition essentielle selon Flichy (2008) pour la stabilisation d'un cadre d'usage.

Ce cadre d'usage ne constitue pas un modèle prescriptif, mais plutôt une grille de lecture et d'action permettant aux institutions de développer des stratégies adaptées à leur contexte spécifique. Sa validité et sa pertinence devront être évaluées dans des contextes disciplinaires variés et à différentes échelles institutionnelles.

# 6. DISCUSSION ET CONCLUSION

Ce travail présente plusieurs contributions originales à la compréhension des processus d'appropriation des IAG dans l'enseignement supérieur. Sur le plan théorique, il enrichit le modèle de Flichy (2008) en précisant le rôle fondamental de l'AE numérique dans la construction du cadre d'usage, tout en l'articulant avec les exigences du cadre réglementaire européen émergent. Sur le plan méthodologique, il propose et valide une échelle d'auto-efficacité numérique spécifiquement adaptée au contexte des IAG et destinée aux enseignants-chercheurs. Sur le plan empirique, il établit des relations statistiquement significatives entre les profils d'AE et les modalités d'appropriation de ces technologies.

La convergence entre nos résultats empiriques et les principes du cadre réglementaire européen est particulièrement notable. Alors que le Règlement IA met l'accent sur l'alphabétisation IA (Article 4) et la supervision humaine (Article 14), nos données démontrent que ces capacités sont précisément modulées par les niveaux d'AE numérique des enseignants. Cette convergence renforce la pertinence de notre approche différenciée, qui permet d'aligner les exigences réglementaires avec les réalités psychologiques et professionnelles du terrain.

Nos résultats font écho à ceux de Mah et Groß (2024) concernant les profils d'enseignants vis-à-vis de l'IA, tout en apportant une dimension supplémentaire par l'articulation avec les niveaux d'AE. La typologie que nous proposons (« Investis, Réservés réflexifs, Réfractaires critiques ») affine la compréhension des dynamiques d'appropriation et permet d'envisager des stratégies d'accompagnement plus ciblées. En particulier, l'absence d'association entre le profil des « Réfractaires critiques » et un niveau particulier d'AE confirme que les résistances à l'adoption des IAG ne se réduisent pas à des questions de compétence technique, mais engagent des considérations épistémologiques et éthiques plus profondes.

Notre étude présente certaines limites qu'il convient de reconnaître. Sa nature transversale ne permet pas d'appréhender les dynamiques d'évolution des profils dans le temps. L'échantillon, bien que diversifié, pourrait ne pas représenter adéquatement certaines disciplines ou institutions spécifiques. Enfin, le caractère déclaratif des pratiques rapportées constitue une limite inhérente à ce type d'approche méthodologique.

Ces limites ouvrent des perspectives de recherche prometteuses : études longitudinales suivant l'évolution des profils au fil du temps, analyses disciplinaires différenciées, observations directes des pratiques, ou encore évaluation de l'impact des dispositifs d'accompagnement inspirés du cadre d'usage proposé. De futures recherches pourraient également explorer les interactions entre les profils d'AE et les catégories de risque des systèmes d'IA, afin d'affiner les stratégies d'accompagnement selon la criticité des applications.

Dans un contexte où les IAG transforment rapidement le paysage académique, notre cadre d'usage fondé sur l'AE fournit des repères conceptuels et pratiques pour une appropriation réfléchie et différenciée de ces technologies (Holmes et al., 2025). Il rappelle que l'intégration des IAG dans l'enseignement supérieur ne saurait se réduire à une question technique ou réglementaire, mais constitue fondamentalement un enjeu de développement professionnel et de construction collective de normes pédagogiques légitimes. A ce titre, il reconnaît l'agentivité individuelle de chaque enseignant-chercheur (Tali Otmani, 2025) dans sa façon d'intégrer ou non les IAG dans ses pratiques professionnelles.

Fatiha TALI OTMANI